\documentclass[12pt]{article}
\usepackage{amsmath, amsthm}

\usepackage{smile}
\usepackage{graphicx,psfrag,epsf}
\usepackage{enumerate}
\usepackage{natbib}
\usepackage{url} % not crucial - just used below for the URL 
\usepackage{zhlipsum}
\usepackage{hyperref}
\hypersetup{
    colorlinks=true,
    citecolor=blue,
    linkcolor=blue,
    urlcolor=magenta
}
\usepackage{bm}
\usepackage{algorithm}
\usepackage{algpseudocode}
\usepackage{booktabs} 
\usepackage[subscriptcorrection]{newtxmath}

\newcommand{\blind}{1}

\usepackage{xr}
\begin{document}

\def\spacingset#1{\renewcommand{\baselinestretch}%
{#1}\small\normalsize} \spacingset{1}

%%%%%%%%%%%%%%%%%%%%%%%%%%%%%%%%%%%%%%%%%%%%%%%%%%%%%%%%%%%%%%%%%%%%%%%%%%%%%%

\if1\blind
{
\title{\bf Transfer Learning for Degree-Corrected Mixed Membership Network Models}
\author{
Yong He$^{*\dagger}$\\
Institute for Financial Studies, Shandong University
\and
Kangxiang Qin$^{\dagger}$\\
Institute for Financial Studies, Shandong University
\and
Haoran Tang$^{\dagger}$\\
School of Mathematics, Shandong University
}
    
\maketitle

\begingroup
\renewcommand{\thefootnote}{}
\footnotetext[0]{\noindent$^{*}$Corresponding author, email: heyong@sdu.edu.cn.}
\footnotetext[0]{\noindent$^{\dagger}$The authors contributed equally to this work.}
\endgroup
} \fi

\if0\blind
{
  \bigskip
  \bigskip
  \bigskip
  \begin{center}
    {\LARGE\bf Transfer Learning for Degree-Corrected Mixed Membership Network Models}
\end{center}
  \medskip
} \fi

\bigskip
\begin{abstract}
Statistical analysis of network data has attracted considerable attention in recent years, due to the rapid advancement of well-trained network models and the  accessibility of large public network datasets. In this article, we propose a transfer learning procedure for boosting estimation accuracy of a target network structure based on the well-known  Degree-Corrected Mixed-Membership (DCMM) model in the literature. By leveraging  useful information from  informative source datasets, we theoretically prove that  the transfer learning procedure greatly improve the estimation accuracy for the target connection probability matrix.  Our
theoretical analysis also reveals  that the benefits from knowledge transfer in this context attributes to the enlarged eigenvalue gap of the target connection probability matrix. Additionally, we propose a random projection step in conjunction with the conventional aggregation procedure to alleviate the heavy computational burden in practice. In the presence of potentially harmful sources, we
further provide an iterative truncation algorithm for selecting useful datasets and avoiding negative transfer. Numerical results showcase the practical utility of our methods in real-world network dataset analysis, including journal citation network dataset and international trade network dataset. 
% The enhancement of estimating the target connection probability matrix may be of independent interest for further in-depth application in, i.e., social networks, economics, and biology.

\end{abstract}

\noindent%

\spacingset{1.45} % DON'T change the spacing!

\section{Introduction}
% With the advent of the big data era, multi-source networks have become increasingly common across various statistical fields. Representative examples include multiple ecological networks among organisms in different regions \citep{Guimera2005} and social networks generated from different platforms, such as Facebook, Twitter and GooglePlus \citep{Zhang2020}. Although these networks are collected from different locations or platforms, they often describe relationships among objects that share similar underlying feature. Thus, these networks is expected to be governed by similar interaction mechanisms, leading to a common latent network structure. Meanwhile, focusing solely on the common structure may be inadequate for multi-network analysis.  Differences among practical 
% conditions can induce systematic deviation from the common network structure. Such deviation is expected to be captured by latent network-specific structure. Thus, taking both network structures into account allows for a more comprehensive characterization of given networks. 

With the advent of the big data era, multi-network datasets have become increasingly prevalent across a wide range of statistical applications. A representative example arises from international trade networks observed over consecutive years \citep{jiang2019227}.  The latent network structure induced by real-world international trade among different nations contains high-value economic information, which motivates a large amount of related research.
In addition to the international trading systems, similar patterns are widely observed in other domains, including article citation networks among various journals \citep{gao2023large} and social networks across multiple digital platforms such as Facebook, Twitter, and GooglePlus \citep{Zhang2020}. 

For each type of network, although individual networks may be collected at different times or in different locations, they often describe relationships among objects with similar underlying characteristics. As a result, networks within the same type are expected to be governed by similar interaction mechanisms, giving rise to a common shared network structure.
At the same time, focusing solely on the shared structure is often inadequate for multi-network analysis. In practice, heterogeneity across networks can induce systematic deviations from the shared structure, which are expected to be captured by a private network structure. Ignoring such heterogeneity may lead to a biased estimation or the loss of network-specific structural information. Therefore, it is essential to jointly model both shared and private structures to achieve a more comprehensive characterization of the observed networks.

However, most existing methods for network analysis are not designed to simultaneously accommodate both shared and network-specific structures. On the one hand, some works \citep{hunter2012computational,mallya2018packnet,jin2025network,jin2025optimal} concentrate on a single network, thereby laying all attention on network-specific structure and completely ignoring the common structural information that may be shared across other similar networks. On the other hand, the research for studying the statistical properties of multi-networks remains insufficient while those existing works  \citep{hansen2005knowledge,mallya2018piggyback} prefer merging all networks into a large network or representing them through a multi-layer network. Although these methods may be effective for capturing shared structure among networks, they often do so at the expense of ignoring their individual private network structures. To fill this gap, in this work we propose a unified statistical  framework for modeling both the shared and network-specific structures of multiple networks simultaneously.

In fact, jointly modeling both shared and individual specific characteristics of a statistical object is a fundamental issue that is ubiquitous and important across a wide range of application fields. For instance, \cite{he2024representational} considered the 
shared and patient-specific feature extraction in gray-scale chest scan image inpainting, while  \cite{shi2024personalized} investigated the identification problem of annually specific key topics among common debating topics in presidential debate transcripts. These studies to some extent underscore the significance of modeling common and network-specific structures in network analysis. Meanwhile, transfer learning, a key branch of modern machine learning that originated in computer science \citep{torrey2010transfer,zhuang2020comprehensive}, serves as a useful tool of enabling accurate estimation of shared features while preserving individual-specific characteristics. This also aligns well with the need to simultaneously capture common and network-specific structures for network analysis. In this work, we adapt the mindset of transfer learning to model both the shared and individual-specific (private) network structures simultaneously for multi-layer network analysis. 

\subsection{Closely related Works}
The purpose of transfer learning is to address unsatisfactory estimation performance on a target dataset by leveraging information from useful source datasets \citep{olivas2009handbook}.
In the supervised learning setting, \cite{Bastani2020} and \cite{Li2021} pioneered a two-step framework for high-dimensional linear regression. This methodology was later extended to generalized linear models \citep{tian2023transfer}, quantile regression models \citep{qiao2023transfer}, and nonparametric classification \citep{Cai2021,reeve2021adaptive}, respectively.
In contrast to supervised transfer learning, progress in unsupervised learning is relatively limited. Principal Component Analysis (PCA) is one of the most widely used unsupervised methods for dimension reduction \citep{pearson1901principal,hotelling1933analysis}. Recently, \cite{DUAN2024105521} first proposed an optimal weighted-covariance-pooling solution called target PCA to study transfer learning for PCA, which performs data integration by using a weighted linear combination of the target covariance matrix and source covariance matrices with  elaborately designed weights. \cite{li2024knowledge} further developed the transfer PCA method and revealed that the enhancement of transfer learning in the context of PCA is attributed to the enlarged eigenvalue gap of the target covariance matrix through a fine-tuning step. Subsequently, \cite{hu2025aggregated} and \cite{he2025transpca} proposed  transfer learning procedures for multi-view datasets in the context of factor models.

% These developments indicate that transfer learning is especially useful in settings where one aims to borrow strength across related datasets while preserving dataset-specific heterogeneity. Multi-source network data fall naturally into this setting, as related networks may share similar structural patterns while still exhibiting network-specific network feature.

A central task in network analysis is the characterization of latent community structure. Existing research can be broadly classified into two categories: the non-overlapping community networks, in which each node belongs to a single community \citep{HOLLAND1983109,abbe2017community}, and the mixed membership networks, which allow nodes to belong to multiple communities \citep{airoldi2008mixed,ball2011efficient,mao2017mixed,mao2018overlapping,mao2021estimating}. Among the latter, the Mixed-Membership Stochastic Block (MMSB) model  proposed by \cite{airoldi2008mixed} is a popular representative model. However, MMSB does not account for degree heterogeneity. To address this limitation, \cite{jin2017estimating} proposed the Degree-Corrected Mixed-Membership (DCMM) model, which accommodates both degree heterogeneity and mixed membership. \cite{jin2024mixed} further introduced the Mixed-SCORE algorithm to efficiently estimate mixed membership vectors for DCMM. \cite{shen2023fadi} further investigated the DCMM model and  estimated the subspace of the connection probability matrix under distributed computation settings. 

Compared to single network analysis, multi-layer networks can benefit from modeling multi-dimensional interactions for real-world networks. This motivates increasing interest in modeling networks with inherently multi-layer structure \citep{wang2021optimal,pensky2019dynamic}. For instance, \cite{pham2021comgcn} used non-adjacent node similarity scores for link prediction in a multi-layer network. \cite{he2025joint} proposed  a novel multi-layer graphon to jointly estimate the edge probabilities for multi-layer networks. \cite{weylandt2025multivariate} stacked multiple networks into semi-symmetric tensor network data and utilized CP decomposition to analyze the network structure. See also \cite{xu2023covariate,jing2021community} for further related discussion about multi-layer network analysis.

\subsection{Main Contributions}
The main contributions  are summarized as follows.
First, we propose a Transfer Learning DCMM (TDCMM) method by modeling each network as a connection probability matrix along with  the corresponding parameter matrices. We also transform the discussion of the network structure into the principal subspace of the connection probability matrix. A shared-private decomposition is also imposed for characterizing both common shared and private structures given multiple networks.
Second, we develop two algorithms that can enhance the corresponding subspace estimations for the connection probability matrix under both oracle (the useful source network datasets are pre-known) and non-oracle (the useful source network datasets are unknown) settings. Moreover, we refine our algorithms by incorporating a random projection procedure, which alleviates the computational burden of processing high-dimensional matrices while incurring minimal information loss.
Third, we establish a comprehensive theoretical framework for the proposed estimators under mild assumptions.
When the source networks are sufficiently informative, the theoretical advantage of our proposed TDCMM methods lies in the enlarged eigenvalue gap of the target connection probability matrix, which originates from the fine-tuning step imposed in our algorithms. This enhancement essentially increases the signal level for estimating the target network structure and the corresponding parameter matrices, thereby yielding superior theoretical performance compared to traditional single DCMM.
Finally, we conduct thorough simulations under different scenarios to show the superiority of TDCMM. Furthermore, we investigate downstream empirical studies on international trade networks \citep{jiang2019227} and journal citation networks \citep{gao2023large}. Results indicate that our proposed   algorithms produce network estimates that are more consistent with real-world patterns and are more interpretable, highlighting the significance of identifying both shared and private network structures in practice.

\subsection{Organization and Notations}
\textit{\textbf{Organization}}: Section \ref{Methodology} introduces the proposed TDCMM framework. Section \ref{algorithm} specifies the design of the oracle and non-oracle TDCMM algorithms. Section \ref{Statistical Theory} develops the statistical theory for the estimators formed by TDCMM algorithms. Section \ref{Numerical Simulations} reports comprehensive simulation studies evaluating the estimation accuracy and robustness of the oracle and non-oracle TDCMM algorithms under different noise levels. Section \ref{Empirical Analysis} presents an empirical application to an international trade network to illustrate the practical usefulness of TDCMM.

\textit{\textbf{Notations}}: We denote random variables by $x,y$. For sequences of random variables $\{x_n\}_{n \geq 1}$ and $\{y_n\}_{n \geq 1}$, we write $x_n \gtrsim y_n$ if $y_n = O_p(x_n)$ and $x_n \lesssim y_n$ if $x_n = O_p(y_n)$; if both hold, we write $x_n \asymp y_n$. The sub-exponential norm of a random variable $x$ is denoted by $\|x\|_{\psi_1}$. Constants are represented by $C$, $c$, $\ldots$, which may vary across different contexts. For a real number $a$, $\lfloor a \rfloor$ denotes the floor function and $\lceil a \rceil$ denotes the ceiling function.

We denote random vectors by bold lowercase letters, e.g., $\ba$, $\bb$. The $i$-th element of a vector $\ba$ is written as $(a)_i$ or $a_i$. $\ba$ and $\bb$ are said to be orthogonal if their inner product is zero, i.e., $\ba^\top \bb = 0$. The $\ell_q$ norm of a vector $\ba$ is denoted by $\|\ba\|_q$, where $q\in[0,\infty]$. For sets $\mathcal{S}$ and $\mathcal{U}$ with $\mathcal{S} \subseteq \mathcal{U}$, the complement of $\mathcal{S}$ is expressed as $\mathcal{S}^c$ or $\mathcal{U} \setminus \mathcal{S} = \mathcal{S}^c \triangleq \{ \bx \in \mathcal{U} : \bx \notin \mathcal{S} \}$. The subspace of $\mathbb{R}^d$ spanned by $\{\bv_1, \bv_2, \ldots, \bv_k\}$ is denoted as $\operatorname{span}(\{\bv_1, \bv_2, \ldots, \bv_k\})$. For a subspace $\mathcal{W}\subseteq\mathbb{R}^d$ with dimension $r < d$, the orthogonal complement is denoted by $\mathcal{W}^\perp$, such that $\mathbb{R}^d = \mathcal{W} \oplus \mathcal{W}^\perp$ and $\bw^\top \bv = 0$ for all $\bw \in \mathcal{W}$ and $\bv \in \mathcal{W}^\perp$.

We denote matrices by bold uppercase letters, e.g., $\Ab$, $\Bb$. The entry of a matrix $\Ab$ at the $i$-th row and $j$-th column is denoted as $(\Ab)_{ij}$ or $\Ab_{ij}$. The Frobenius norm of $\Ab$ is written as $\left\|\Ab\right\|_F$, and the spectral norm is denoted by $\left\|\Ab\right\|_2$ or $\sigma_{\max}(\Ab)$. The trace of $\Ab$ is expressed as $\operatorname{tr}(\Ab)$. A diagonal matrix with entries $a_1, a_2, \ldots, a_n$ on the diagonal is written as $\operatorname{diag}(a_1, a_2, \ldots, a_n)$. The eigenvalues of a symmetric matrix $\Ab\in\mathbb{R}^{p\times p}$ are denoted as $\{\lambda_i(\Ab)\}_{i=1}^{\text{rank}(\Ab)}$. If $\Ab$ is non-symmetric, $\{\sigma_i(\Ab)\}_{i=1}^{\text{rank}(\Ab)}$ denotes the corresponding singular values. A matrix $\bXi \in \mathbb{R}^{d \times r}$ is called column orthogonal if it satisfies $\bXi^\top \bXi = \Ib_r$. The projection matrix onto a subspace spanned by column orthogonal matrix $\bXi$ is denoted as $\boldsymbol{P} = \bXi \bXi^\top$, where $\bXi$ is column orthogonal.

For real sequences $\{a_n\}_{n \geq 1}$ and $\{b_n\}_{n \geq 1}$, we write $a_n \gtrsim b_n$ if there exists a constant $C > 0$ such that $a_n \geq C b_n$ for all $n \geq 1$; similarly, $a_n \lesssim b_n$ if $a_n \leq C b_n$ for all $n \geq 1$; and $a_n \asymp b_n$ if both hold.

\section{Methodology}
\label{Methodology}
In this section, we start by reviewing the traditional DCMM framework in Section \ref{traditional DCMM}. Then, we further provide a detailed introduction to our TDCMM model in Section \ref{TDCMM}.

\subsection{Review of DCMM}
\label{traditional DCMM}
Suppose we have an undirected network $\mathcal{G}$ with $d$ nodes and a $K$-community structure.
The DCMM model proposed by \cite{jin2017estimating} describes $\mathcal{G}$ in three parts. $(1)$ DCMM imposes a degree heterogeneity parameter $\theta_i > 0$ for node $i$ in order to describe the activity level of each node $i$. We collect them in a diagonal matrix $\bTheta = \operatorname{diag}(\theta_1, \ldots, \theta_d)$. $(2)$ DCMM also assigns each node $i$ a mixed membership vector $\bpi_i = ((\pi_i)_1, \ldots, (\pi_i)_K)^\top$ for characterizing the belonging relationship for $i$ and each community, where $(\pi_i)_k$ denotes the probability that node $i$ belongs to the $k$-th community. $(3)$ DCMM imposes a symmetric, non-negative matrix $\Pb \in \mathbb{R}^{K \times K}$ to depict the connectivity between each pair of communities, where $\Pb_{k_1 k_2}$ represents the probability that the $k_1$-th community and the $k_2$-th community connect. 
Given the above three parameter matrices, DCMM generates a network adjacency matrix $\Xb$ according to the following model:
$$
\Xb_{ij} \sim \operatorname{Bernoulli}(\Hb_{ij}), \ \text{where}\  \Hb = \bTheta \bPi \Pb \bPi^\top \bTheta^\top\ \text{and}\ \bPi = (\bpi_1, \ldots, \bpi_d)^\top.
$$
Equivalently, we can rewrite this probability as $\mathbb{P}(\Xb_{ij} = 1) = \theta_i \theta_j \bpi_i^\top \Pb \bpi_j$. Node $i$ is termed a pure node for the $k$-th community if $(\pi_i)_k = 1$. Otherwise, it is a mixed node. Given $\bTheta$, $\bPi$ and $\Pb$, we can fully characterize $\mathcal{G}$. Accordingly, the DCMM enables us to transform the issue of fully recovering network $\mathcal{G}$ into estimating $\bTheta$, $\bPi$ and $\Pb$ based on the observed $\Xb$.

\cite{jin2024mixed} proposed the Mixed-SCORE algorithm to estimate $\bTheta$, $\bPi$ and $\Pb$. To be specific, let $\Hb = \bXi \bLambda \bXi^{\top}$ and $\Xb = \hat{\bXi}\hat{\bLambda}\hat{\bXi}^\top$ be the eigenvalue decompositions of $\Hb$ and $\Xb$ respectively, where $\bXi = (\bxi_1, \ldots, \bxi_K)$ and $\widehat{\bXi} = (\widehat{\bxi}_1, \ldots, \widehat{\bxi}_K)$. Define $\Rb = (\br_1, \br_2, \ldots, \br_d)^{\top} \in \mathbb{R}^{d \times (K-1)}$ as
$$
\Rb_{ik} = (\bxi_{k+1})_i / (\bxi_{1})_{i}, \quad 1 \le i \le d,\, 1 \le k \le K-1.
$$
Under the conditions in Lemma 2.1 of \cite{jin2024mixed}, $\{\br_i\}_{i=1}^d$ lie within a $(K-1)$-dimensional simplex $\mathcal{S} \subset \mathbb{R}^{K-1}$, and $\br_i$ is a vertex of $\mathcal{S}$ if and only if node $i$ is a pure node.
TThus, \cite{jin2024mixed} proposed the Mixed-SCORE algorithm. In detail, we first estimate $\widehat{\Rb}=(\widehat{\br}_1, \widehat{\br}_2, \ldots, \widehat{\br}_d)^{\top}$ with entries $\hat{\Rb}_{ik} = (\hat{\bxi}_{k+1})_i / (\hat{\bxi}_{1})_{i}$, and then utilize $\hat{\Rb}$ to acquire $\hat{\bPi}=(\hat{\bpi}_1,\ldots,\hat{\bpi}_K)$ and the $K$ vertices $\hat{\Vb}=(\hat{\bv}_1,\ldots,\hat{\bv}_K)$ of the point cloud $\{\widehat{\br}_i\}_{i=1}^d$.
Then we proceed to construct
$$(\hat{\bb}_1)_k=[\lambda_1(\Xb)-\hat{\bv}_k^\top \operatorname{diag}(\lambda_2(\Xb),\cdots,\lambda_K(\Xb))\hat{\bv}_k]^{-1/2}\ \text{and}\ \hat{\Bb}=\operatorname{diag}(\hat{\bb}_1) (\Ib_K, \hat{\Vb}^\top),$$
and the estimators of $\Pb$ and $\bTheta$:
\begin{equation*}
\begin{matrix}
\widehat{\Pb} = \widehat{\Bb} \widehat{\bLambda} \widehat{\Bb}^\top,\ \text{and}\ 
\widehat{\bTheta}_{ii}
=\widehat{\theta}_i = (\widehat{\bxi}_1)_i / (\widehat{\bpi}^\top_i \widehat{\bb}_1), \quad 1 \leq i \leq d.
\end{matrix}
\end{equation*}
Combining $\hat{\Pb}$, $\hat{\bTheta}$ and $\hat{\bPi}$ gives us
$
\widehat{\Hb} = \widehat{\bTheta} \widehat{\bPi} \widehat{\Pb} \widehat{\bPi}^\top \widehat{\bTheta}^\top.
$ 
We conclude the details of the Mixed-SCORE algorithm in Algorithm $2$ in supplement for saving space.

% \begin{lemma}[Lemma 2.2 of \cite{jin2024mixed}]
% \label{lem:P_Theta}
% Let $\bLambda = \operatorname{diag}(\lambda_1(\Hb), \lambda_2(\Hb), \cdots, \lambda_K(\Hb))$ and $\Vb = (\bv_1, \bv_2, \cdots, \bv_K)$ be the vertices of the simplex $\mathcal{S}$. Define $(\bb_1)_k = [\lambda_1(\Hb)-\bv_k^\top \operatorname{diag}(\lambda_2(\Hb),\cdots,\lambda_K(\Hb))\bv_k]^{-1/2}$, $\Bb = \operatorname{diag}(\bb_1) (\Ib_K, \Vb^\top)$. Then, under the conditions in Lemma 2.1 of \cite{jin2024mixed}, we have
% \begin{align*}
% \Pb = \Bb \bLambda \Bb^\top,\ \text{and}\ 
% \theta_i = (\bxi_1)_i / (\bpi^\top_i \bb_1), \quad 1 \leq i \leq d.
% \end{align*}
% \end{lemma}

% By estimating $(\bb_1)_k$ as $(\hat{\bb}_1)_k=[\lambda_1(\Xb)-\hat{\bv}_k^\top \operatorname{diag}(\lambda_2(\Xb),\cdots,\lambda_K(\Xb))\hat{\bv}_k]^{-1/2}$, we can obtain  $\hat{\Bb}=\operatorname{diag}(\hat{\bb}_1) (\Ib_K, \hat{\Vb}^\top)$.
% According to Lemma \ref{lem:P_Theta}, the estimation of $\Pb$ and $\bTheta$ can be acquired as
% \begin{equation*}
% \begin{matrix}
% \widehat{\Pb} = \widehat{\Bb} \widehat{\bLambda} \widehat{\Bb}^\top,\ \text{and}\ 
% \widehat{\theta}_i = (\widehat{\bxi}_1)_i / (\widehat{\bpi}^\top_i \widehat{\bb}_1), \quad 1 \leq i \leq d,
% \end{matrix}
% \end{equation*}

\subsection{Transfer Learning for DCMM}
\label{TDCMM}

Consider a collection of $ M $ undirected networks $\{\mathcal{G}_m\}_{m=1}^M$, where the networks share a common node set $ \mathcal{V}$ with $|\mathcal{V}|=d$ while each $\mathcal{G}_m$ has $ K_m $ communities. We let $ \mathcal{G}_1 $ be the target network of primary interest, and $ \mathcal{G}_2, \dots, \mathcal{G}_M $ serve as source networks that may provide useful structural information. For each $m\in[M]:=\{1,2,\ldots,M\}$, we model each $\mathcal{G}_m$ using DCMM:
\begin{align}\label{TDCMM_model}
    \mathbb{P}((\Xb_m)_{ij}=1) = (\Hb_m)_{ij} = \theta_i^m \theta_j^m (\bpi_i^m)^\top \Pb^m \bpi_j^m,
\end{align}
where $ \Hb_m = \bTheta^m \bPi^m \Pb^m (\bPi^m)^\top (\bTheta^m)^\top $.
In order to fully recover $\Hb_1$, we only need to focus on estimating $\bPi^1$, $\bTheta^1$ and $\Pb^1$ by leveraging the network structural information from $\{\Hb_m\}_{m=2}^M$. 
Denote the leading $K_m$-dimensional eigenvectors of $\Hb_m$ as $\bXi_m = (\bxi_1^m, \ldots, \bxi_{K_m}^m) $; then our TDCMM framework starts with a shared-private decomposition for each $\bXi_m$. To be specific, we assume that each $\mathrm{span}(\bXi_{m})$ can be decomposed into two orthogonal subspaces as follows:
\begin{equation}
\label{eq:span(p+s)}
\mathrm{span}(\bXi_m) = \mathrm{span}(\bXi_m^s) \oplus \mathrm{span}(\bXi_m^p),
% \ (\bXi_{m}^{s})^\top\bXi_{m}^{s}=\Ib_{K_s},\ (\bXi_{m}^{p})^\top\bXi_{m}^{p}=\Ib_{K_m-K_s}.
\end{equation}
where $\bXi_k^{s}\in\mathbb{R}^{d\times K_s}$ and $\bXi_k^{p}\in\mathbb{R}^{d\times (K_m-K_s)}$ are column orthogonal matrices and we term $\mathrm{span}(\bXi_{m}^{s})$ and $\mathrm{span}(\bXi_{m}^{p})$ as the shared and private subspaces for $\Hb_m$ respectively. Let $\bXi_{m}\bXi_{m}^{\top}$, $\bXi_{m}^{p}(\bXi_{m}^{p})^{\top}$, and $\bXi_{m}^{s}(\bXi_{m}^{s})^{\top}$ be the orthogonal projection matrices of the corresponding subspaces, and we could also write
\eqref{eq:span(p+s)} in a projection matrix form:
\begin{equation}\label{eq:p+s}
\bXi_{m}\bXi_{m}^{\top} =  \bXi_{m}^{p}(\bXi_{m}^{p})^{\top} + \bXi_{m}^{s}(\bXi_{m}^{s})^{\top}, \  \text{where} \  \bXi_{m}^{s}(\bXi_{m}^{s})^{\top}\bXi_{m}^{p}(\bXi_{m}^{p})^{\top} = \textbf{0}_{d\times d}.
\end{equation}

The reasons for making such a decomposition for each $\bXi_m$ lie in the following two aspects.

First, \eqref{eq:p+s} allows us to provide a latent common-specific network structure decomposition in a mathematical form.  
For each $m\in[M]$, $\mathrm{span}(\bXi_m)$ represents the eigenspace of $\Hb_m$ while $\Hb_m$ fully depicts $\mathcal{G}_m$. Thus, $\bXi_m(\bXi_m)^\top$ can characterize the latent network structure of $\mathcal{G}_m$. As we intend to jointly model the transferable common structure across $\{\mathcal{G}_m\}_m$ and the network-specific structure, it is natural to further decompose  $\bXi_m\bXi_m^\top$ into the corresponding shared part, i.e. $\bXi_m^s(\bXi_m^s)^\top$, and the private  (network-specific) part, i.e. $\bXi_m^p(\bXi_m^p)^\top$. With such  decomposition in hand, we are able to capture both shared and network-specific structures in a more principled manner, providing a more  comprehensive characterization of $\mathcal{G}_1$.

Second, \eqref{eq:p+s} can not only characterize both shared and private network structures, but also improve the subsequent parameter matrices estimation for $\Hb_1$. 
As outlined in Section \ref{traditional DCMM}, a key step in estimating parameter matrices is the construction of the point cloud matrix $ \hat{\Rb}^1 $ via $ \hat{\bXi}_1 = (\hat{\bxi}_1^1, \ldots, \hat{\bxi}_{K_1}^1) $.
Thus, the estimation accuracy of $ \hat{\bXi}_1 $ directly affects the performance of $ \widehat{\Rb}^1 $ and  the parameter matrices. The traditional DCMM in \cite{jin2017estimating} relies solely on $ \Xb_1 $ and thus often yields limited estimation precision. In contrast, with \eqref{eq:p+s} in hand, we can  transfer the useful network structural information shared among $\{\Xb_m\}_{m=2}^M$ to $\Xb_1$ and thereby enhance the estimation accuracy of $\hat{\bXi}_1^s$. Then, the remaining task is to apply $\Xb_1$ to estimate $\bXi_1^p$, which is expected to be easier than directly estimating $\bXi_1$. Thus, we can obtain improved estimates for $\bXi_1$ and the downstream parameter matrices compared to the traditional DCMM. 

For the convenience of  discussion, we denote $\cI \subseteq [M] \backslash \{1\}$ as the index set of the informative networks that contain transferrable structural information. Furthermore, the ``closeness" between every two shared network structures also requires formal definition. As  the shared network structure is represented by $\mathrm{span}(\bXi_m^s)$, we use the sin$\Theta$ metric to measure this closeness and assume the following similarity condition:
\begin{align}
\label{eq:information level characterization}
\left\| \bXi_{m}^{s}(\bXi_{m}^{s})^{\top} - \bXi_{1}^{s}(\bXi_{1}^{s})^{\top} \right\|_F \le h,\ \text{for}\ m\in\cI.
\end{align}
Intuitively, \eqref{eq:information level characterization} requires the informative source networks to be close enough to the target one from a subspace angle-based perspective. In particular, \eqref{eq:information level characterization} can be transformed into the following inequality:
$$
\sqrt{2}\left\|(\bXi_m^s)^{\top}(\bXi_1^s)^{\perp}\right\|_F\leq h.
$$
If we denote the singular values of $(\bXi_m^s)^{\top}(\bXi_1^s)^{\perp}$ as $\{\sigma_{m,i}\}_{i=1}^{K_s}$, then the principal angles between $\text{span}((\bXi_1^s)^{\perp})$ and $\text{span}(\bXi_m^s)$ are defined as $\{\zeta_{m,i}\triangleq\cos^{-1}(\sigma_{m,i})\}_{i=1}^{K_s}$. As $h$ tends to $0$, $\zeta_{m,i}\approx\pi/2$ holds for each $i\in[K_{s}]$. Thus, $\text{span}(\bXi_m^{s})$ and $\text{span}((\bXi_1^{s})^{\perp})$ are almost orthogonal, implying a close distance between $\text{span}(\bXi_m^{s})$ and $\text{span}(\bXi_1^{s})$. For those non-informative networks where $m\notin\cI$, their ``common" network structure is not suitable for transferring to $\mathcal{G}_1$, i.e, the angle between $\text{span}(\bXi_m^s)$ and $\text{span}((\bXi_1^s)^{\perp})$ can be arbitrarily small, leading to a large distance between $\text{span}(\bXi_m^s)$ and $\text{span}(\bXi_1^s)$.

\section{Algorithms}
\label{algorithm}

In light of \eqref{eq:p+s}, the main task for our TDCMM algorithms reduces to obtaining well-estimated $\bXi_1^s$ and $\bXi_1^p$.
The basic framework of TDCMM algorithms contains three parts in particular. First, we construct a detailed procedure for obtaining $\tilde{\bXi}_s$ by exploiting the useful source network structure under the scenarios when $\cI$ is either known or unknown. We then include a fine-tuning step to obtain the estimate of $\bXi_1^p$ based on the well-estimated $\tilde{\bXi}_s$. Combining these two steps leads to the third part, where we obtain the final estimators of $\bPi^1$, $\bTheta^1$, and $\Pb^1$. The following two sections specify these three steps in detail.

\subsection{Oracle TDCMM Algorithm}
\label{Oracle TDCMM Algorithm}
In this section, we assume that $\cI = [M] \backslash \{1\}$ is given in advance. 
For the first part of the oracle TDCMM, $\bXi_1^s$ can be estimated by leveraging the following optimization \citep{fan2019distributed,shi2024personalized}:
\begin{align}
\label{eq:GB}
\widehat{\bXi}_s\widehat{\bXi}_s^\top = \argmax_{\bXi \in \mathcal{O}(d,K_s)} \frac{1}{M} \sum_{m \in \{1\} \cup \cI} \mathrm{tr}(\widehat{\bXi}_m\widehat{\bXi}_m^\top \bXi\bXi^\top),
\end{align}
where $\mathcal{O}(d,K_s)\triangleq\{\bXi\in\mathbb{R}^{d\times K_s}|\bXi^\top\bXi = \Ib_{K_s}\}$ and $\widehat{\bXi}_m$ is the estimate of $\bXi_m$ obtained from the leading $K_m$ eigenvectors of $\Xb_m$.
Normally, \eqref{eq:GB} can be solved by the leading $K_s$ eigenvectors of the weighted average projection matrix $\widehat{\bSigma}$:
\begin{equation}\label{eq:aveproj}
    \widehat{\bSigma} = \frac{1}{M}\sum_{m\in\{1\}\cup\cI} \widehat{\bXi}_m\widehat{\bXi}_m^\top.
\end{equation}
However, we consider a more challenging scenario where the server that stores $\widehat{\bSigma}$ faces a high computational requirement. Under this circumstance, a traditional SVD of $\widehat{\bSigma}$ requires $O(d^3)$ flops, which may impose a severe computational burden on the corresponding server.

\begin{figure}
\begin{center}
\includegraphics[width=6.5in]{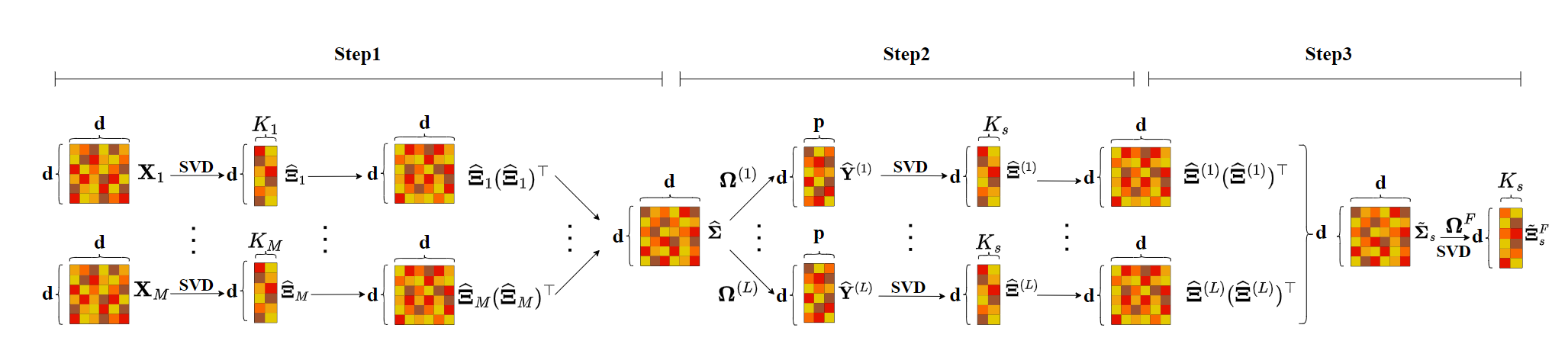}
\end{center}
\caption{\label{Fast Transfer PCA} Schematic diagram of the shared subspace estimation steps in Oracle FTPCA.}
\end{figure}

To alleviate the computational burden of SVD for $\hat{\bSigma}$, we impose a two-step method by utilizing the random projection mindset \citep{halko2011finding,chen2016integrating,shen2023fadi}. Suppose  $\min\{p,p'\}>K_s$ while $q>1$, we draw $L$ independent random matrices $\{\bOmega^{(l)}\}_{l \in [L]}$ where each $\bOmega^{(l)}\in \mathbb{R}^{d \times p}$ has independent standard Gaussian entries. For the first step, we compute $\{\widehat{\Yb}^{(l)} = \widehat{\bSigma} \bOmega^{(l)}\}_{l\in[L]}$ and their corresponding top-$K_s$ eigenvectors $\{\widehat{\bXi}^{(l)}\}_{l\in[L]}$. We further acquire the corresponding average projection estimators $\tilde{\bSigma}_s = \sum_{l = 1}^L \widehat{\bXi}^{(l)}(\widehat{\bXi}^{(l)})^\top/L$ for reducing the potential information loss caused by a single random projection. For the second step, we again define $\bOmega^F \in \mathbb{R}^{d \times p'}$ with independent standard Gaussian entries and compute $\tilde{\bXi}_s^F$ \footnote{For better comprehension, $F$ in the superscript stands for ``Fast".} by the leading $K_s$ left eigenvectors of the matrix $\tilde{\Yb}_{s}^F = (\tilde{\bSigma}_{s})^q \bOmega^F$ where $q$ denotes the number of power iterations. These two steps  are summarized as steps $2$ and $3$ in Figure \ref{Fast Transfer PCA} for clear demonstration.

Generally, the computational cost  for the first step scales as $O(Ldp^2)$ while that of the second step is of order $O(dK_mp'Lq)$.  Compared to the traditional SVD, whose computational cost is of order $O(d^3)$, our random projection mindset owns a significant computational efficiency enhancement as long as $d \gg p' \geq \max(2K_s, K_s + 7)$, $p \geq \max(2K_s, K_s + 8q - 1)$ and $p^2L\leq d^2$.

The second part of the oracle TDCMM pertains to acquiring the estimator of  $\bXi_1^p$. This estimator is obtained by transferring $\tilde{\bXi}_s^F$ back to $\Xb_1$ and performing a fine-tuning step. Specifically, we compute the leading $(K_1 - K_s)$ eigenvectors of the following projected adjacency matrix
$$
\Xb_1^p = \left[\Ib_p - \tilde{\bXi}_s^F(\tilde{\bXi}_s^F)^\top\right] \Xb_1 \left[\Ib_p - \tilde{\bXi}_s^F(\tilde{\bXi}_s^F)^\top\right],
$$
to obtain the private subspace estimator $\tilde{\bXi}_p^F$. 
One may consider the following special but intuitive case for explaining the theoretical gain of the second fine-tuning step. We assume that $\Hb_1$ has the following eigenvalues
$$\underbrace{\lambda^{p}_1(\Hb_1) \geq \cdots \geq\lambda^{p}_{K_1-K_s}(\Hb_1)}_{\text{network-specific}}\geq \underbrace{\lambda^{s}_{K_1-K_s+1}(\Hb_1)\geq \cdots\geq\lambda^{s}_{K_1}(\Hb_1)}_{\text{common shared}}>\lambda_{K_1+1}(\Hb_1)\geq \cdots\geq \lambda_{p}(\Hb_1),$$
where the eigenvalues correspond to private and shared subspaces, respectively. If we perform SVD with dimension $K_1$ using only the $\Xb_1$, then based on the Davis-Kahan theorem, the performance of traditional DCMM depends on the eigenvalue gap $\lambda^{s}_{K_1}(\Hb_1)-\lambda_{K_1+1}(\Hb_1)$. On the other hand, if we firstly seek an accurate estimate of $\bXi_1^s$, i.e. $\tilde{\bXi}_s^F$, by borrowing latent network structural information from $\{\Xb_m\}_{m\in\cI}$ and deflating it from $\Xb_1$, which results in $\Xb_1^p$. Then, we can expect that only the private eigenvalues are retained in the second step and the eigenvalue gap of $\Xb_1^p$ becomes $\lambda^{p}_{K_1-K_s}(\Hb_1)-\lambda_{K_1+1}(\Hb_1)$, which is larger than the traditional eigenvalue gap $\lambda^{s}_{K_1}(\Hb_1)-\lambda_{K_1+1}(\Hb_1)$. 
See Algorithm \ref{alg:Oracle FTPCA Algorithm} for further details of the first two parts of oracle TDCMM. Note that the partitioning of $\{1\} \cup \cI$ in Algorithm \ref{alg:Oracle FTPCA Algorithm}
is introduced solely for the subsequent theoretical analysis and is not needed in practice.

\begin{algorithm}
\caption{The First Two Parts of Oracle TDCMM}
\label{alg:Oracle FTPCA Algorithm}
\begin{algorithmic}[1]
\Require Target and informative adjacency matrices with their subspace dimension $\{\Xb_m, K_m\}_{m\in\{1\}\cup\cI}$, shared subspace dimension $K_s$, number of random projection matrices $L$, random projection dimension parameter $p$;
\Ensure 
\State Divide $\{1\}\cup\cI$ into $\cI_1\cup\cI_2$ where $\cI_1\cap\cI_2=\emptyset$ and $|\cI_1|=|\cI_2|$ (if not, let $|\cI_1|=|\cI_2|+1$);
\Statex \textbf{Shared Subspace Estimation:}
\For{each $m \in \cI_1$}
    \State Acquire $\widehat{\bXi}_m$ by computing the top $K_m$ left singular vectors of $\Xb_m$;
\EndFor
\State Aggregate all $\widehat{\bXi}_m$ to obtain $\widehat{\bSigma} = \sum_{m\in\cI_1}\widehat{\bXi}_m\widehat{\bXi}_m^\top/|\cI_1|$;
\State Generate i.i.d. standard Gaussian matrices $\{\bOmega^{(l)}\}_{l=1}^L$;
\For{each $l \in [L]$}
    \State Acquire $\widehat{\bXi}^{(l)}$ by the top $K_s$ left singular vectors of $\widehat{\Yb}^{(l)} = \widehat{\bSigma}\bOmega^{(l)}$;
\EndFor
\State Compute $\tilde{\bSigma}_s=\sum_{l=1}^L \widehat{\bXi}^{(l)}\big(\widehat{\bXi}^{(l)}\big)^\top/L$, and obtain $\tilde{\bXi}_s^{(1)}$ as the leading $K_s$ left singular vectors of $\tilde{\bSigma}_s$;
\State Generate i.i.d. $\{\bOmega^{(l)}\}_{l=L+1}^{2L}$;
\State Apply $\{\bOmega^{(l)}\}_{l=L+1}^{2L}$ and $\{\Xb_m\}_{m\in\cI_2}$ to acquire $\tilde{\bXi}_s^{(2)}$ by shared subspace estimation process;

\Statex \textbf{Private Subspace Estimation:}
\State Form $\Xb_1^p = (\Ib_p-\tilde{\bXi}_s^{(1)}(\tilde{\bXi}_s^{(1)})^\top)\Xb_1(\Ib_p-\tilde{\bXi}_s^{(2)}(\tilde{\bXi}_s^{(2)})^\top)$;
\State Compute the top $(K_1 - K_s)$ left singular vectors, $\tilde{\bXi}_p$, of $\Xb_1^p$;

\State \textbf{Return} $\tilde{\bXi}_1^F = (\tilde{\bXi}_p^F, \tilde{\bXi}_s^F)\triangleq(\tilde{\bXi}_p^F, \tilde{\bXi}_s^{(1)})$.
\end{algorithmic}
\end{algorithm}

For the third part, we use $\tilde{\bXi}_1^F$ as a plug-in estimator in the Mixed-SCORE Algorithm to estimate the remaining three parameter matrices of DCMM. Ultimately, we yield an estimate of the target probability matrix $\Hb_1$ by the oracle TDCMM procedure. The  process for the last part is concluded in Algorithm \ref{alg:TDCMM}. Oracle TDCMM is motivated by many practical applications. Consider $d$ specific users as the nodes. Their social activities datasets on $M$ different social platforms can be transformed into adjacency matrices $\Xb_1,\ldots,\Xb_M$. Since the group of users is nearly fixed across most of the platforms, $\Xb_1,\ldots,\Xb_M$ are expected to share a latent common network structure. We can then utilize this advantage to improve the estimation of the target probability matrix $\Hb_1$ while maintaining its own network-specific structural information, which leads to a more accurate description of the network structure for this group of users.

\begin{algorithm}
\caption{The Third Part of Oracle TDCMM}
\label{alg:TDCMM}
\begin{algorithmic}[1]
\Require $\tilde{\bXi}_1^F=\left(\widehat{\bxi}_{1}^{1,F},\widehat{\bxi}_{2}^{1,F},\ldots,\widehat{\bxi}_{K_1}^{1,F}\right)$, $K_1$, adjacency matrix $\Xb_1$, threshold $T = \log(d)$;
\Ensure

\State Plug $(\Xb_1,\tilde{\bXi}_1^F,K_1,T)$ into Mixed-SCORE algorithm to get $\widehat{\bPi}^1$, $\widehat{\Vb}^1 = (\widehat{\bv}_1^1,\ldots,\widehat{\bv}_{K_1}^1)$ and $\widehat{\bb}^1_1$;
\State Form $\widehat{\Bb}^1 = \text{diag}(\widehat{\bb}^1_1)(\textbf{1}_{K_1},(\widehat{\Vb}^1)^\top)$;
\State Compute $\widehat{\Pb}^1 = \widehat{\Bb}^1 \widehat{\bLambda}^1 (\widehat{\Bb}^1)^\top$;
\State Compute $\widehat{\theta}_i^1 = (\widehat{\bxi}_{1}^{1,F})_{i}/(\widehat{\bpi}_i^1)^\top \widehat{\bb}^1_1$ for $i=1,\ldots,d$;
\State Set $\widehat{\bTheta}^1 = \text{diag}(\widehat{\theta}_1^1,\ldots,\widehat{\theta}_d^1)$;
\State \textbf{Return} $\widehat{\Hb}_1 = \widehat{\bTheta}^1 \widehat{\bPi}^1 \widehat{\Pb}^1 (\widehat{\bPi}^1)^\top (\widehat{\bTheta}^1)^\top$.
\end{algorithmic}
\end{algorithm}

\subsection{Non-oracle TDCMM Algorithm}

In fact, $\cI$ is largely unknown in practice. Therefore, there is a chance that one may also include some non-informative networks in the study, which requires us to develop a non-oracle TDCMM framework to automatically estimate $\cI$. To achieve this purpose, we transform \eqref{eq:GB} into the following ``truncated'' optimization:
\begin{equation}\label{eq:non-oracle}
\hat{\bXi}^F_{s,\tau}(\hat{\bXi}^F_{s,\tau})^{\top}=\argmax_{\bXi\in \mathcal{O}(d,K_s)}\frac{1}{M}\operatorname{tr}\left(\hat{\bXi}_1\hat{\bXi}_1^\top\bXi\bXi^\top\right)+ \sum_{m\in [M]\backslash\{1\}}\frac{1}{M}\max\left\{\operatorname{tr}\left(\widehat{\bXi}_m\widehat{\bXi}_m^\top\bXi\bXi^\top\right),K_s-\tau\right\},
\end{equation}
The main difference between \eqref{eq:non-oracle} and \eqref{eq:GB} lies in $\tau$. Different values of $\tau$ correspond to different types of optimization tasks. If $\tau \geq K_s$, solving \eqref{eq:non-oracle} is equivalent to blindly pooling all $\{\widehat{\bXi}_{m}\}_{\in [M]}$ and performing Algorithms \ref{alg:Oracle FTPCA Algorithm} and \ref{alg:TDCMM}, thereby having no network selection ability. If $\tau \leq 0$, all $\{\widehat{\bXi}_m\}_{m=2}^M$ are filtered out by the threshold, and the solution of \eqref{eq:non-oracle} depends only on the individual PCA estimator $\hat{\bXi}_1$ computed from the target dataset. If $\tau\in (0,K_s)$, then \eqref{eq:non-oracle} is a non-convex optimization problem with network selection capability, and $\tau$ controls the strength of network selection.

Once the non-oracle subspace estimator $\tilde{\bXi}^F_{s,\tau}$ is obtained, the subsequent steps follow exactly the same as  Algorithm \ref{alg:Oracle FTPCA Algorithm}. Thus, the remaining task reduces to solving problem \eqref{eq:non-oracle}. In fact, one can simply use manifold gradient-descent-type (first-order) or Newton-type (second-order) algorithms to iteratively update $\tilde{\bXi}_{s,t-1}^F$ to $\tilde{\bXi}_{s,t}^F$, given a suitable initial estimator \citep{huper2004newton,helmke2007newton,shi2024personalized}.
However, the local convergence for manifold iterative algorithms may fail if a poor initial estimator is provided. Thus, to strengthen numerical robustness, we instead apply an iterative truncated approach. Suppose that we have access to $\tilde{\bXi}_{s,t-1}^F$ and $\cI_{t-1}$ at the beginning of the $t$-th iteration.
Our iterative truncated approach adopts a  mindset that is similar to K-means. We view $\{\widehat{\bXi}_m\}_{m\in\cI_{t-1}}$ as ``nodes'' in $\mathcal{O}(d,K_s)$ and then compute the distance between each $\hat{\bXi}_m$ and $\hat{\bXi}_1$ using the distance metric imposed in \eqref{eq:non-oracle}. We end the $t$-th iteration step by filtering out the useless $\hat{\bXi}_m$ and update $\cI_{t-1}$ into
\begin{equation*}
\label{eq:I_t}
    \cI_{t}=\left\{m\in\cI_{t-1}\mid \operatorname{tr}\left(\widehat{\bXi}_m\widehat{\bXi}_m^\top\tilde{\bXi}_{s,t-1}^F(\tilde{\bXi}_{s,t-1}^F)^\top\right)\geq K_s-\tau \right\}.
\end{equation*}

Owing to the initialization-robustness of the K-means algorithm, a precise initial estimate is not necessary for our iterative procedure. Meanwhile, applying K-means mindset helps us identify  those $\{\hat{\bXi}_m\}_m$ that are close to $\hat{\bXi}_1$. These two advantages allow us to mitigate the risk of negative transfer and alleviate heavy dependence on the initial estimate. Besides, the computational complexity of our truncated iteration approach is within an acceptable level. The data selection parameter $\tau$ in Algorithm \ref{alg:non-oracle FTPCA} is selected by an additional cross-validation step. See Algorithm \ref{alg:non-oracle FTPCA} for details of non-oracle TDCMM.

\section{Statistical Theory}
\label{Statistical Theory}

% In this section, we develop the theoretical framework of TDCMM. Without loss of generality, we estimate the shared signal subspace $\tilde{\bXi}_s$ using the pooled index set $\{1\}\cup I$, which replaces the estimator $\tilde{\bXi}_s^{(1)}$. We first introduce a set of assumptions that support the subsequent analysis. We then establish theoretical results for the subspace estimator under the oracle TDCMM. Under mild conditions, the oracle TDCMM can outperform the traditional DCMM. Building on the convergence rate of the oracle TDCMM, we further study the statistical properties of $\widehat{\Rb}^1$, $\widehat{\Pb}^1$, $\widehat{\bPi}^1$, and $\widehat{\bTheta}^1$, which serve as key components for downstream estimation of the DCMM parameter matrices. We conclude with the non-oracle TDCMM and show that the optimization problem in \eqref{eq:non-oracle} admits a local maximum property.

In this section, we present a thorough theoretical discussion for the TDCMM framework. 
We begin by introducing a set of assumptions that underpin the subsequent analysis. Then we establish theoretical guarantees for the subspace estimator from the oracle TDCMM. We show that, under mild conditions, the theoretical guarantee for $\tilde{\bXi}_1^F$  outperforms the traditional DCMM. Based on the improved target subspace estimator, we further derive statistical properties of $\widehat{\Rb}^1$, $\widehat{\Pb}^1$, $\widehat{\bPi}^1$, and $\widehat{\bTheta}^1$. We conclude by studying the non-oracle TDCMM and show that the optimization problem \eqref{eq:non-oracle} satisfies a local maximum property.

\begin{algorithm}[H]
\caption{The First Two Parts of Non-oracle TDCMM}
\label{alg:non-oracle FTPCA}
\begin{algorithmic}[1]
\Require Shared dimension $K_s$, $\{\Xb_m, K_m\}_{m\in[M]}$, threshold parameter $\tau$, number of iterations $T$, initial shared subspace $\tilde{\bXi}_{s,0}^F$, initial source set $\cI_0 = [M]\backslash\{1\}$;
\Ensure
\State Compute $\{\widehat{\bXi}_m\}_{m\in[M]}$ from $\{\Xb_m\}_{m\in[M]}$;
\For{$t = 1$ to $T$}
    \State Acquire $\cI_t = \left\{ m \in \cI_{t-1} \mid \operatorname{tr}\left[\tilde{\bXi}_{s,t-1}^F\left(\tilde{\bXi}_{s,t-1}^F\right)^\top \widehat{\bXi}_m(\widehat{\bXi}_m)^\top\right] \geq K_s -  \tau \right\}$;
    \State Apply Algorithm \ref{alg:Oracle FTPCA Algorithm} to $\{\Xb_m\}_{m \in  \{1\}\cup \cI_t}$ to obtain $\tilde{\bXi}_{s,t}^F$;
\EndFor
\State Run Private Subspace Estimation step of Algorithm \ref{alg:Oracle FTPCA Algorithm} on $\{\Xb_m\}_{m \in \{1\}\cup\cI_T}$ and $\tilde{\bXi}_{s,T}^{F}$ to obtain $\tilde{\bXi}_{p,T}^{F}$;
\State \textbf{Return} $\tilde{\bXi}_{1,\tau}^{F} = (\tilde{\bXi}_{p,T}^{F}, \tilde{\bXi}_{s,T}^{F})$.
\end{algorithmic}
\end{algorithm}

\subsection{Basic Assumptions}
\begin{assumption}
\hypertarget{ass:theta_order}{}
\label{ass:theta_order}
There exist constants $C_1, C_2 > 0$ such that, for all $m \in \{1\}\cup\cI$,
\begin{align*}
C_1 \theta_{\max}^1 \leq \theta^m_{\max} \leq C_2 \theta_{\max}^1, \quad
C_1 \|\btheta^1\|_1 \leq \|\btheta^m\|_1 \leq C_2 \|\btheta^1\|_1.
\end{align*}
where $\left\|\btheta^m\right\|_1 = |\theta^m_{1}| + |\theta^m_{2}| + \cdots + |\theta^m_{d}|$ and $\theta_{\max}^m = \max(\theta_1^m,\cdots,\theta_d^m)$,
\end{assumption}

\begin{assumption}
\hypertarget{ass:Convergence of E}{}
\label{ass:Convergence of E}
For each $m \in \{1\}\cup\cI$,
let $\Eb_m = \Xb_m - \Hb_m$ be the error matrix. Assume that $\Eb_m$ follows a sub-exponential distribution, and there exists a rate $r(\btheta)$ such that
\begin{align*}
\left\| \left\|\Eb_m\right\|_2 \right\|_{\psi_1} = \sup_{q \geq 1} q^{-1} (\mathbb{E} \left\|\Eb_m\right\|_2^q)^{1/q} \lesssim r_m(\btheta) \lesssim r(\btheta).
\end{align*}
\end{assumption}

% \begin{remark}
% Assumption \ref{ass:Convergence of E} is adopted from \cite{jin2024mixed} and is common in network-related articles such as \cite{jin2017estimating,jin2024mixed,shen2023fadi}. Based on Assumption \ref{ass:Convergence of E} and \cite{vershynin2010introduction}, there exists a constant $c_e > 0$ such that, for any $t > 0$, $$\mathbb{P}(\left\|\Eb_m\right\|_2 \geq t\bigr) \leq \exp\bigl(-c_e t / r(\btheta)\bigr),$$ which gives us  $\left\|\Eb_m\right\|_2 = O_p\bigl(r(\btheta)\bigr)$.
% \end{remark}
% \begin{remark}
% Assumption \ref{ass:Convergence of E} is adopted from \cite{jin2024mixed} and is widely used in the network literature, such as \cite{jin2017estimating,jin2024mixed,shen2023fadi}. It imposes a sub-exponential tail condition on the spectral norm of the noise matrix. By standard concentration results (e.g., \cite{vershynin2010introduction}), there exists a constant $c_e > 0$ such that, for any $t > 0$,
% $$
% \mathbb{P}(\|\Eb_m\|_2 \geq t) \leq \exp\bigl(-c_e t / r(\btheta)\bigr).
% $$
% As a result, we have $\|\Eb_m\|_2 = O_p\bigl(r(\btheta)\bigr)$. This condition is mild and ensures that the noise level is well controlled by $r(\btheta)$.
% \end{remark}
% To ensure the statistical performance of the shared subspace in the first step, we impose the following similarity assumption.

\begin{assumption}
\label{ass:Identifiability}
According to \eqref{eq:p+s}, we assume that the shared subspace of each informative source satisfies
$$
\left\|\bXi_m^s(\bXi_m^{s})^\top - \bXi_1^s(\bXi_1^{s})^\top\right\|_F \leq h\ \text{for}\ m\in\cI.
$$
For the private component, we assume that there exists a sequence of positive constants $\{g_M\}_{M\geq 2}$ with $g_M \to 0$ as $|\cI| \to \infty$ such that
$$
\left\| \frac{1}{|\cI|+1} \sum_{m\in\{1\}\cup\cI} \bXi_m^p(\bXi_m^{p})^\top \right\|_2 \leq g_M.
$$
\end{assumption}

% \begin{remark}
% Assumption \ref{ass:Identifiability} mainly requires: $(1)$ the shared subspaces are close enough to each other from a subspace-metric perspective; and $(2)$ the private subspaces are well separated so that their aggregation generates no large eigenvalues that could compromise the identification of the shared subspace.
% % \end{remark}
% \begin{remark}
% Assumption \ref{ass:Identifiability} contains two parts. First, it requires that the shared subspaces across different sources are close under the subspace distance, which guarantees that a common structure can be reliably estimated. Second, it requires that the private subspaces are sufficiently dispersed, so that their average does not produce large eigenvalues. This separation prevents the private components from interfering with the identification of the shared subspace. Both conditions are essential for ensuring identifiability.
% \end{remark}

\begin{assumption}
\hypertarget{ass:eigenvalue separation}{}
\label{ass:eigenvalue separation}
Let the parameters satisfy $\tilde{\beta} = o(1)$ and $\beta = o(\tilde{\beta})$, and assume that:
\begin{itemize}
    \item The leading eigenvector of $\Hb_1$ is identifiable, i.e., $\lambda_1(\Hb_1) - \lambda_2(\Hb_1) \asymp \|\btheta^1\|_2^2 \cdot {K_1}^{-1}$.
    \item Define $\Hb_1^p = (\Ib_d-\bXi_1^s(\bXi_1^s)^\top) \Hb_1 (\Ib_d-\bXi_1^s(\bXi_1^s)^\top)$ and $d_p(\Hb_1^p) \triangleq d_{K_1 - K_s}(\Hb_1^p) = \lambda_{K_1 - K_s}(\Hb_1^p) - \lambda_{K_1 + 1}(\Hb_1^p)$, such that
    $d_p(\Hb_1^p) \asymp \tilde{\beta} \cdot \|\btheta^1\|_2^2 \cdot {K_1}^{-1}$.
    \item The ${K_1}$-th eigenvalue gap of $\Hb_1$ satisfies $\Delta \triangleq |\lambda_{K_1}(\Hb_1)| = \beta \cdot \|\btheta^1\|_2^2 \cdot {K_1}^{-1}$.
    \item $\lambda_1(\Hb_1)-\left|\lambda_2(\Hb_1)\right| \asymp\lambda_1(\Hb_1)$, and $\left|\lambda_k(\Hb_1)\right| \lesssim \tilde{\beta}K_1^{-1}\|\btheta^1\|_2^2$ for $2 \leq k \leq K_1$.
\end{itemize}
\end{assumption}

The above assumptions are standard in the analysis of multi-network models. Assumptions \ref{ass:theta_order} and \ref{ass:Convergence of E} are adapted from \cite{jin2024mixed} and are commonly used in the network literature. They impose a sub-exponential tail condition on the spectral norm of the noise matrix, which implies $\|\Eb_m\|_2 = O_p\bigl(r(\btheta)\bigr)$ and ensures that the noise level is well controlled. Assumption \ref{ass:Identifiability} requires that the shared subspaces across informative sources are close to each other, while the private subspaces are sufficiently dispersed, so that the common structure can be identified without substantial interference from source-specific components. Assumption \ref{ass:eigenvalue separation} is also adapted from \cite{jin2024mixed}, with the additional condition $\beta = o(\tilde{\beta})$. This condition ensures that, after removing the shared component, the effective eigen-gap in the projected matrix $\Hb_1^p$ is larger than the original boundary level $\Delta$, which is the key reason why the oracle TDCMM can achieve improved performance. Since the projected matrix mainly reflects the private signal, this requirement is mild and is often satisfied when the signal and noise are reasonably separated.
% \end{remark}

\subsection{Oracle TDCMM}

The convergence rate of the oracle TDCMM estimator is formalized in the following theorem.

\begin{theorem}
\hypertarget{Thm:oracle convergence rate}{}
\label{Thm:oracle convergence rate}
(Oracle TDCMM)
Define
\begin{equation*}
\mathbf{S} = \sqrt{\frac{K_sd}{L|\cI|p}} \frac{r(\btheta)}{\Delta} + \sqrt{\frac{K_s}{|\cI|}}\frac{r(\btheta)}{\Delta}  + \sqrt{K_s}\exp\left( -\frac{c}{8} \sqrt{\frac{p}{d}}  R^{-1} \right)+ \sqrt{\frac{K_sd}{Lp}} h,
\end{equation*}
where $c$ is a constant and
\begin{equation*}
R = \sqrt{\frac{K_s}{|\cI|}} \frac{r(\btheta)}{\Delta} + \sqrt{K_s}\frac{r(\btheta)^2}{\Delta^2} + h.
\end{equation*}
% Under Assumptions \ref{ass:theta_order} to \ref{ass:eigenvalue separation}, and let $d \gg p \geq \max(2K_s,K_s+7)$, there exists a constant $c > 0$ such that:
% \begin{align}
% \label{eq:oracle convergence rate}
%     \mathbb{E}\left\|\bXi_1\bXi_1^\top - \tilde{\bXi}_1\tilde{\bXi}_1^\top\right\|_F \lesssim \frac{r(\btheta)}{d_p(\Hb_1^p)} +  \mathbf{S}.
% \end{align}
% Furthermore
If both $d \gg p' \geq \max(2K_s,K_s+7)$ and $ p \geq \max(2K_s,K_s+8q-1)$ hold, then based on Assumptions \ref{ass:theta_order} to \ref{ass:eigenvalue separation}, the following convergence rate holds:
\begin{align}
\label{eq:fast oracle convergence rate}
    \mathbb{E}\left\|\bXi_1\bXi_1^\top - \tilde{\bXi}_1^F(\tilde{\bXi}_1^F)^\top\right\|_F \lesssim \frac{r(\btheta)}{d_p(\Hb_1^p)} +  \mathbf{S} + \mathbf{F},
\end{align}
where $\mathbf{F} = \sqrt{\frac{K_s d}{p}} \left( 2\sqrt{\frac{d}{p'}} R \right)^q$, $r(\btheta) \asymp \sqrt{\theta_{\max}^1 \left\|\btheta^1\right\|_1}$and $q$ is defined in step $3$ in Figure \ref{Fast Transfer PCA}.
\end{theorem}

As shown in \eqref{eq:fast oracle convergence rate}, the convergence rate exhibits a phase-transition behavior \citep{li2022transfer}. The upper bound for the estimation error of $\tilde{\bXi}_1^F$ consists of a private-subspace term $r(\btheta)/d_p(\Hb_1^p)$ and a shared-subspace term $\mathbf{S}+\mathbf{F}$. When $|\cI|$ is small, the transfer gain is mainly reflected by the decrease of the shared-subspace error as $|\cI|$ increases. When $|\cI|$ is sufficiently large and $h$ is sufficiently small, $\mathbf{S}+\mathbf{F}$ becomes negligible, and the overall rate is dominated by $r(\btheta)/d_p(\Hb_1^p)$. Compared with the corresponding convergence rate for the traditional DCMM in \cite{jin2024mixed}, the improvement of oracle TDCMM comes from the fact that the leading term is controlled by $d_p(\Hb_1^p)$ rather than the original boundary eigen-gap $\Delta$. Hence, when $d_p(\Hb_1^p)\gg \Delta$, oracle TDCMM achieves a strictly faster rate. The reason is that TDCMM first uses source networks to estimate the shared subspace and then removes this part through the fine-tuning step, so that the projected matrix $\Hb_1^p$ mainly retains the private signal. As a result, the effective eigen-gap for subspace estimation is enlarged from $\Delta$ to $d_p(\Hb_1^p)$. Assumption \ref{ass:eigenvalue separation} formalizes this feature through the condition $\beta=o(\tilde{\beta})$, which is mild under our framework because, after the shared component is removed, the projected matrix is expected to have a clearer private signal-noise separation. Therefore, the faster rate of oracle TDCMM is ultimately due to the enlarged effective eigen-gap induced by the shared-private decomposition and the fine-tuning step, which improves upon the corresponding error bound in \cite{jin2024mixed}. Moreover, the additional term $\mathbf{F}$ only reflects the information loss caused by random projection, and it is negligible when $q \asymp \log(d)$ is chosen properly.

Based on Theorem \ref{Thm:oracle convergence rate}, we provide the error bound for estimating $\Pb^1,\bPi^1$ and $\bTheta^1$. To derive explicit parameter bounds, we impose the following additional assumptions.

\begin{assumption}
\label{ass:θ上界的假设}
Let $\theta_{\min}^m = \min(\theta_1^m,\cdots,\theta_d^m) \quad  m \in \{1\}\cup\cI$. 
We assume that  $\btheta^1$ satisfies $\theta_{\max}^1 \leq C$, where $C>0$ is a constant. Define the following two error terms:
$$
err^* \triangleq \frac{\sqrt{\theta_{\max}^1 \left\|\btheta^1\right\|_1}}{\left\|\btheta^1\right\|_2^2} \cdot \frac{\left\|\btheta^1\right\|_2}{(\theta_{\min}^1)^2\sqrt{d}},\ \text{and}\ 
err \triangleq \frac{\sqrt{\theta_{\max}^1 \left\|\btheta^1\right\|_1}}{\left\|\btheta^1\right\|_2^2} \cdot \frac{\left\|\btheta^1\right\|_2}{(\theta_{\min}^1)^2\sqrt{\log d}}.
$$
We assume that $err\to 0$.
\end{assumption}

\begin{assumption}
\label{ass:P,G上界假设}
Assume that $\Pb^1$ and 
$\Gb = K_1\left\|\btheta^1\right\|_2^{-2}((\bPi^1)^\top(\bTheta^1)^2\bPi^1)$
satisfy $$\max\{\|\Pb^1\|_{\max}, \|\Gb\|_{2}, \|\Gb^{-1}\|_{2}\}\leq C,$$ 
where $C>0$ is a constant.
\end{assumption}

\begin{assumption}
\label{ass:P,G特征值假设}
Assume that the singular values of $\Pb^1\Gb$ satisfy:
\begin{enumerate}
    \item The second largest singular value obeys $|\sigma_2(\Pb^1\Gb)| \leq (1-c_1)\sigma_1(\Pb^1\Gb)$, where $c_1 \in (0,1)$;
    \item There exist a constant $c_1>0$ and a sequence $\beta_n \in (0,1)$ such that
          $$
          c_1\beta_n \leq |\sigma_{K_1}(\Pb^1\Gb)| \leq |\sigma_{2}(\Pb^1\Gb)| \leq c_1^{-1}\beta_n.
          $$
\end{enumerate}
\end{assumption}

\begin{assumption}
\label{ass:P,G的特征向量假设}
Let $\bm{\eta}_k(\Pb^1\Gb)$ be the $k$-th right singular vector of $\Pb^1\Gb$. We assume that:
\begin{enumerate}
    \item For any $1 \leq k \leq K_s$, we have $(\bm{\eta}_1(\Pb^1\Gb))_k > 0$;
    \item The ratio between the largest and smallest components satisfies
          $$
          \frac{\max_{1 \leq k \leq K_s} (\bm{\eta}_1(\Pb^1\Gb))_k}{\min_{1 \leq k \leq K_s} (\bm{\eta}_1(\Pb^1\Gb))_k} \leq C,
          $$
          where $C>0$ is a constant.
\end{enumerate}
\end{assumption}

% Under the above assumptions, we obtain the following two corollaries.
Assumptions \ref{ass:θ上界的假设} to \ref{ass:P,G的特征向量假设} are still adopted from \cite{jin2024mixed}. Under these assumptions, and following a similar proof strategy as \cite{jin2024mixed}, we obtain the following two corollaries.

\begin{corollary}
\label{Coro:R bound}
Based on Assumptions \ref{ass:theta_order} to \ref{ass:P,G的特征向量假设}, the following bound for $\{\widehat{\br}_i^1\}_{i=1}^d$ holds
\begin{align}
    \frac{1}{d} \sum_{i=1}^{d} \left\|\Ab\widehat{\br}_i^1 - \br_i^1\right\|_2^2 = O_p\left(\max\left(  \frac{K_1^3}{\tilde{\beta}^2} err^{*2}, \frac{K_1}{d} \frac{\left\|\theta^1\right\|^2_2}{(\theta_{\min}^1)^2}\mathbf{S},  \frac{K_1}{d} \frac{\left\|\theta^1\right\|^2_2}{(\theta_{\min}^1)^2}\mathbf{F}  \right)\right).
\end{align}
When $M$ and $q$ are sufficiently large, it further holds that
\begin{align}
\label{eq:comparison with DCMM}
    \frac{1}{d} \sum_{i=1}^{d} \left\|\Ab\widehat{\br}_i^1 - \br_i^1\right\|_2^2 = O_p\left(\frac{K_1^3}{\tilde{\beta}^2} (err^*)^{2}\right).
\end{align}
\end{corollary}

% The convergence rate for $\widehat{\Rb}$ is $K_1^3 err^{*2}/\beta^2$ in \cite{jin2024mixed}, which is clearly a low order term compared to \eqref{eq:comparison with DCMM} thanks to Assumption \ref{ass:eigenvalue separation}. Thus, following the same procedure as in \cite{jin2024mixed}, the convergence rate for estimating $\Rb^1$ can be improved under transfer learning.Similarly, for $\Pb^1$, $\bPi^1$, and $\bTheta^1$, we have the following corollary.

\begin{corollary}
\label{Coro:P Pi Theta bound}
When the conditions in Corollary \ref{Coro:R bound} are all satisfied, the estimated parameter matrices $\widehat{\Pb}^1, \widehat{\bPi}^1 $ and $\widehat{\bTheta}^1$ own the following bounds:
\begin{enumerate}
    \item $\left\|\widehat{\Pb}^1 - \Pb^1\right\|_2 =  O_p\left(\bigl(K_1^{2} + K_1^{\frac{5}{2}}\tilde{\beta}^{-1}\bigr)\left\|\btheta^1\right\|_2^{-3}err^*\right);$
    \item $d^{-1}\sum_{i=1}^d\left\|\widehat{\bpi}_i^1 - \bpi_i^1 \right\|_2^2= O_p\left(K_1^3\tilde{\beta}^{-2}(err^*)^2\right);$
    \item $\left\|\widehat{\bTheta}^1 - \bTheta^1\right\|_F^2 = O_p\left(\left\|\btheta^1\right\|_2^{2}K_1^3(\tilde{\beta}')^2 err^2\right).$
\end{enumerate}
Where $\tilde{\beta} = \tilde{\beta}'\sqrt{\log d}$.
\end{corollary}

Compared with the DCMM result in \cite{jin2024mixed}, the main gain in Corollaries \ref{Coro:R bound} and \ref{Coro:P Pi Theta bound} is that the key eigenvalue-gap term in the denominator is enlarged from $\Delta$ to $d_p(\Hb_1^p)$. As a result, the leading term in the estimation error of $\Rb^1$ improves from order $O_p\bigl(K_1^3\beta^{-2}(err^*)^2\bigr)$ in \cite{jin2024mixed} to $O_p\bigl(K_1^3\tilde{\beta}^{-2}(err^*)^2\bigr)$ here, and the same replacement from $\beta$ to $\tilde{\beta}$ also yields sharper rates for estimating $\Pb^1$ and $\bPi^1$. Moreover, under the additional condition $\sqrt{d/(\log d)^2}\beta=o(\tilde{\beta}')$, the convergence rate for estimating $\bTheta^1$ is improved as well. Therefore, compared with \cite{jin2024mixed}, oracle TDCMM achieves better convergence rates for all parameter matrices in \eqref{TDCMM_model}. This improvement comes from the fine-tuning step in TDCMM. After borrowing information from informative source networks to estimate the shared structure, the target network only needs to estimate the remaining private part, which leads to a larger effective eigengap in the second step. Hence, the stronger rates in Corollaries \ref{Coro:R bound} and \ref{Coro:P Pi Theta bound} are ultimately driven by the eigengap enlargement brought by the shared-private decomposition and the transfer learning step.

\subsection{Non-oracle TDCMM}

In this section, we prove that the oracle shared subspace estimator from \eqref{eq:GB} is also a local maximum of \eqref{eq:non-oracle} with high probability. We begin with Assumption \ref{ass:Identifiability2}:

\begin{assumption}
\hypertarget{ass:Identifiability2}{}
\label{ass:Identifiability2}
For $m \in \cI^c$, let $d_m = K_s - \mathrm{tr}[\bXi_m\bXi_m^\top\bXi_1^s(\bXi_1^{s})^\top]$. Assume there exists a constant $d_{\tau} > 0$ such that $d_m > d_\tau$ holds for all $m \in \cI^c$.
\end{assumption}

For $k\in\cI$, we have $0\leq d_m\lesssim h^2$. Informally, $d_{\tau}$ quantifies the smallest deviation between $\bXi_m$ and $\bXi_1^{s}$ when $k\in \cI^c$. 
Based on Assumption \ref{ass:Identifiability2}, we obtain the following theorem:

\begin{theorem}
\hypertarget{thm:Non-oracle TDCMM}{}
\label{thm:Non-oracle TDCMM}
Under Assumptions \ref{ass:theta_order} to \ref{ass:Identifiability2}, if $\mathbf{S} + \max_{m \in \cI^c}(r_m(\btheta)/\Delta_m) = o(d_{\tau})$ as $d \to \infty$ and $h \to 0$, then for any $c \in (0.5,1)$ and $\tau \in [(1-c)d_\tau, c d_\tau]$, the oracle shared subspace estimator $\tilde{\bXi}_s$ from \eqref{eq:GB} is a local maximizer of \eqref{eq:non-oracle} with probability tending to $1$.
\end{theorem}

According to Theorem \ref{thm:Non-oracle TDCMM}, the quantity $d_{\tau}$ measures the minimal separation between $\bXi_m$ and $\bXi_1^s$ for each $m \in \mathcal{I}^c$. The existence of the interval $[(1-c)d_\tau, c d_\tau]$ determines the flexibility and robustness in choosing the threshold $\tau$ for Algorithm \ref{alg:non-oracle FTPCA} in practical applications. Specifically, $d_{\tau}$ typically appears as a nonzero constant and remains of constant order in practice. As a result, the admissible range for $\tau$ also has constant-order length, which provides sufficient freedom in tuning this parameter. In addition, $\tau$ does not need to be specified with high precision to ensure correct identification of $\mathcal{I}$ from an empirical perspective. This insensitivity to the exact choice of $\tau$ enhances the practical usefulness and stability of the proposed non-oracle TDCMM method in complex data applications.

\section{Numerical Simulations}
\label{Numerical Simulations}

In this section, we conduct numerical simulations to examine the performance of the TDCMM framework proposed in Sections \ref{Methodology} and \ref{algorithm}. We divide the source networks into two groups: source networks that share a similar common subspace with the target network, and source networks generated under three scenarios with different noise levels. In Scenario S1, the source networks contain only small perturbations. In Scenario S2, the source networks are subject to moderate contamination. In Scenario S3, the source networks are heavily contaminated, where another unrelated common subspace center is generated independently. The detailed data generation process and the parameter settings are collected in Appendix B in supplement.

\begin{figure}
\begin{center}
\includegraphics[width=5in]{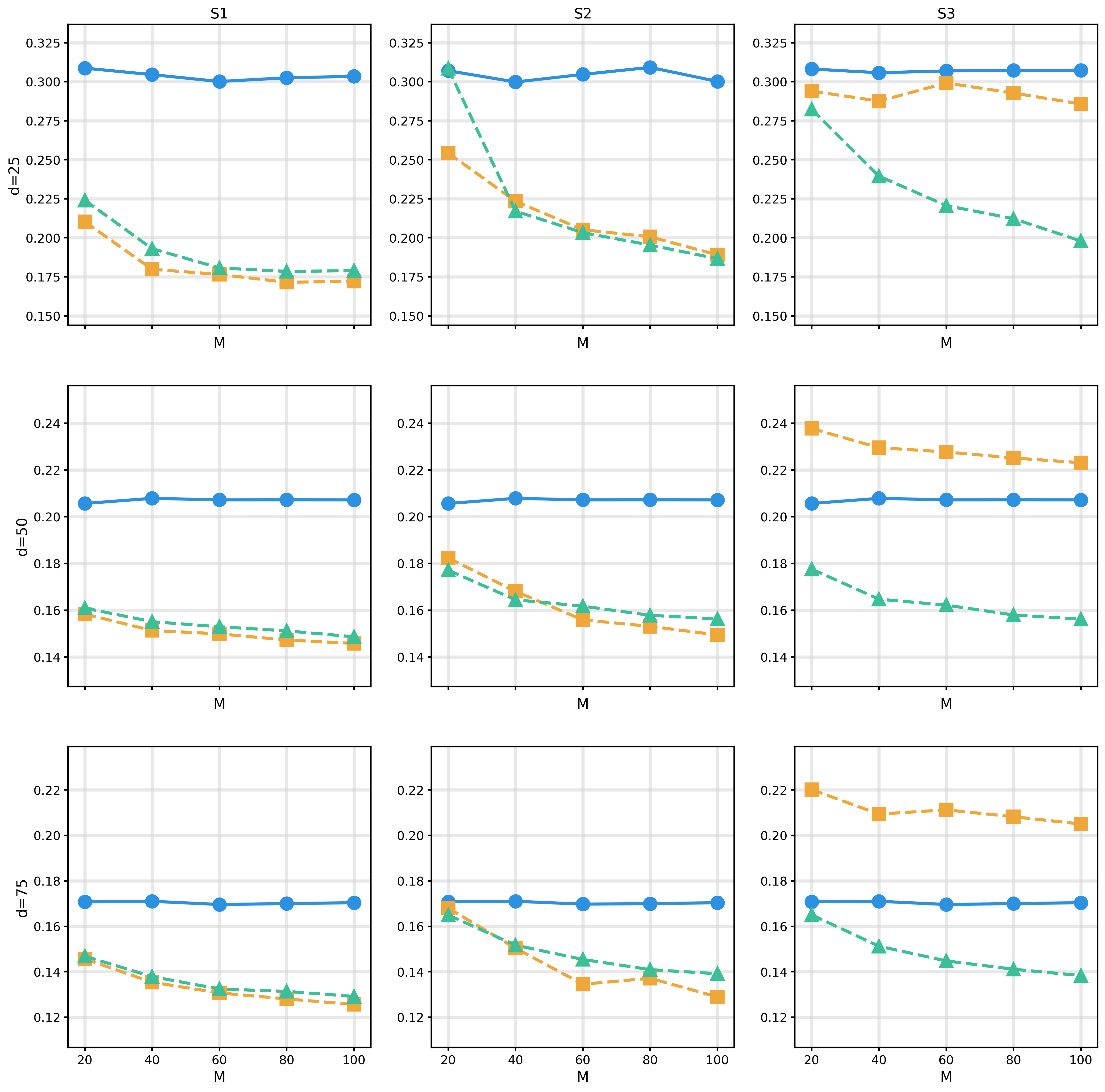}
\end{center}
\caption{D-metric of the probability matrix estimation for three methods under three scenarios with $d = 25,50,75$. The blue solid curve, yellow dashed curve, and green dashed curve correspond to the traditional DCMM, oracle TDCMM, and non-oracle TDCMM, respectively. From top to bottom, each row corresponds to a different value of $d$. From left to right, columns represent results for S1, S2, and S3, respectively. In each subgraph, the $x$-axis represents the number of adjacency matrices, and the $y$-axis represents the estimation error.}
\label{fig:H_error_plot}
\end{figure}

We study the estimation error of $\Hb_1$ under $M = 20,40,60,80,100$. We use the following D-metric to evaluate the estimator $\widehat{\Hb}_1$:
\begin{equation}
\label{eq:D-metric}
\text{error}_H = \frac{\left\|\widehat{\Hb}_1 - \Hb_1\right\|_F}{d}.
\end{equation}
Each setting is repeated $1000$ times, and the average D-metric error is reported in Figure~\ref{fig:H_error_plot}.

In Scenario S1, where all source adjacency matrices are informative, $\text{error}_H$ decreases quickly as $M$ increases. This shows that both oracle and non-oracle TDCMM outperform the traditional DCMM. A similar pattern is observed for oracle TDCMM in Scenario S2, which suggests that it remains effective under mild contamination. In Scenario S3, both the traditional DCMM and oracle TDCMM fail to provide stable estimates because the source networks contain severe noise. By contrast, non-oracle TDCMM still performs better than the other two methods, due to its data selection step. These results show that the TDCMM framework improves estimation accuracy over the traditional DCMM and that modeling both common and network-specific structures is important in multi-network analysis.

\section{Real Data Analysis}
\label{Empirical Analysis}

In this section, we compare the proposed non-oracle TDCMM with the traditional DCMM on an international trade network \citep{jiang2019227}. The parameter settings used in this section are given in Appendix B.

Specifically, \cite{jiang2019227} uses global bilateral merchandise trade data from the UN Comtrade database and constructs a panel of trade networks for 227 countries and regions from 1985 to 2015. In this network, each node represents a country and each directed edge represents a trade flow. There is a directed edge from country $i$ to country $j$ if country $i$ exports goods to country $j$. The trade network can therefore be described by a weighted adjacency matrix, where each entry reflects the volume of trade.

\begin{figure}
\begin{center}
\includegraphics[width=5in]{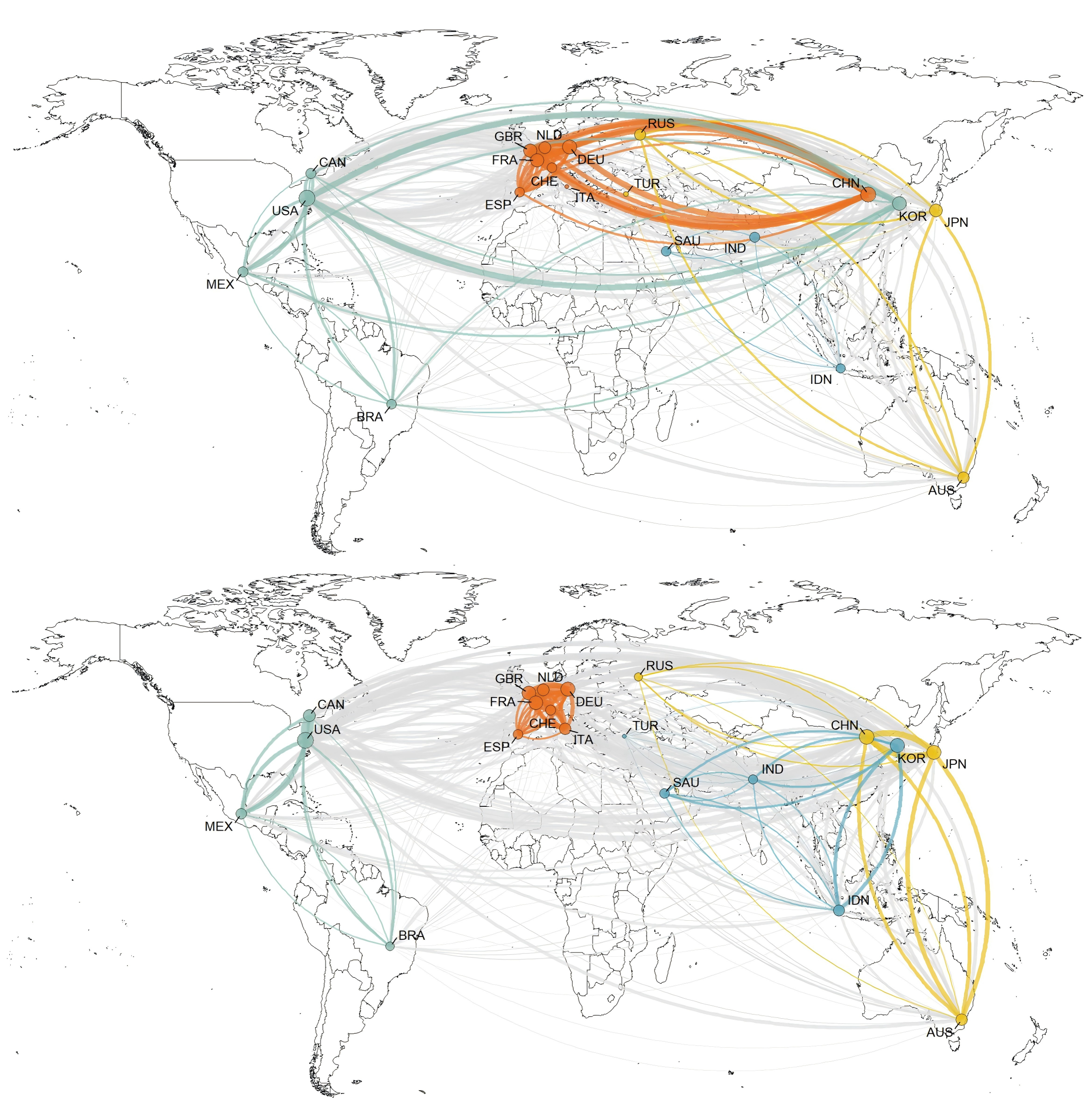}
\end{center}
\caption{The figure displays the estimated international trade network. The top panel shows the community structure obtained from the traditional DCMM, while the bottom panel shows the corresponding results from TDCMM. Different colors indicate different communities identified by TDCMM and DCMM. The size of each node reflects the estimated degree heterogeneity parameter $\theta_i$, and the presence and thickness of edges are determined by the entries of the connection probability matrix $\Hb_{ij}$.}
\label{fig:trade DCMM result}
\end{figure}

To adapt these data to the TDCMM framework, we select the top $20$ countries ranked by gross domestic product in 2024 as nodes. We define an edge between two countries if both bilateral export and import volumes rank among the top $10$. Based on this rule, we construct one adjacency matrix for each year. We take the 2015 network as the target network and use the networks in 1985, 1995, 2005, and 2009 as source datasets. We repeat both methods $100$ times and report the average estimated parameter matrices. The main results are shown in Figure \ref{fig:trade DCMM result}.

Figure \ref{fig:trade DCMM result} shows that edge thickness reflects trade strength between countries, while node size reflects the activity level of each country in the trade network. The results suggest that non-oracle TDCMM gives a more coherent description of the global trade network and is more consistent with real trade patterns. In particular, non-oracle TDCMM identifies four main trade blocks, corresponding to the Americas, Europe, East Asia, and Central and South Asia. By contrast, the traditional DCMM treats the trade network as a single whole and does not provide a clear description of regional structure. This comparison further shows the value of modeling both common and network-specific structures in real data analysis. Additional empirical results for the journal citation network are given in Appendix C.2.

\section{Conclusions and Future Work}

In this paper, we study the estimation of a target network under the DCMM framework from a transfer learning perspective. We propose the TDCMM model to jointly characterize the shared and private structures across multiple networks. We also impose two TDCMM  algorithms. The estimators provided by them are superior in both theoretical and practical perspectives. At last, there are several natural directions for future research of TDCMM including the extensions to Laplacian-based methods, directed network and partially observed network systems.

% Several directions deserve further study. One is to extend the current adjacency-matrix-based framework to Laplacian-based estimation in transfer learning settings. Another is to generalize the proposed method to more complex network data, such as directed, weighted, or dynamic networks. It is also of interest to develop more adaptive source-selection and weighting strategies to better handle heterogeneous source networks and reduce negative transfer. Finally, future work may investigate the use of TDCMM in downstream tasks such as link prediction and missing-edge recovery. 

\bibliographystyle{agsm}

\bibliography{Bibliography-MM-MC}

\end{document}